\documentclass[pra,aps,oneocolumn,nopacs,superscriptaddress,nofootinbib]{revtex4}
\usepackage{graphicx} 
\usepackage{dcolumn}  
\usepackage{bm}       
\usepackage{epsfig}
\usepackage[bbgreekl]{mathbbol}
\usepackage{enumerate}
\usepackage{amsmath}
\usepackage{amsfonts}
\usepackage[T1]{fontenc}
\usepackage[latin1]{inputenc}
\usepackage{lmodern}
\usepackage{amssymb}
\usepackage{pst-all}
\usepackage{psfrag}

\def\ii{{\rm i}}  \def\ee{{\rm e}}  
\def\Rb{{\bf R}}      
    \def\Eb{{\bf E}}  \def\rb{{\bf r}}
\def\xx{\hat{\bf x}}    
\def\th{\vec{\bf{\theta}}}
\def\fE{\vec{\mathcal{E}}}  \def\Mb{{\bf M}}  \def\db{{\bf d}}
\def\neff{{n_{\rm eff}}}   \def\neffdos{{n_{\rm eff}^2}}
\def\omegap{{\omega_{\rm bulk}}}   \def\omegapp{{\omega_{\rm bulk}^2}}

\begin{document}

\title{Plasmonics in Atomically Thin Materials}
\author{F.~Javier~Garc\'{\i}a~de~Abajo}
\email[Corresponding author: ]{javier.garciadeabajo@icfo.es}
\affiliation{ICFO-Institut de Ciencies Fotoniques, Mediterranean Technology Park, 08860 Castelldefels (Barcelona), Spain}
\affiliation{ICREA-Instituci\'o Catalana de Recerca i Estudis Avan\c{c}ats, Passeig Llu\'{\i}s Companys, 23, 08010 Barcelona, Spain}
\author{Alejandro~Manjavacas}
\affiliation{Department of Physics and Astronomy and Laboratory for Nanophotonics, Rice University, Houston, Texas 77005, United States}

\begin{abstract}
The observation and electrical manipulation of infrared surface plasmons in graphene have triggered a search for similar photonic capabilities in other atomically thin materials that enable electrical modulation of light at visible and near-infrared frequencies, as well as strong interaction with optical quantum emitters. Here, we present a simple analytical description of the optical response of such kinds of structures, which we exploit to investigate their application to light modulation and quantum optics. Specifically, we show that plasmons in one-atom-thick noble-metal layers can be used both to produce complete tunable optical absorption and to reach the strong-coupling regime in the interaction with neighboring quantum emitters. Our methods are applicable to any plasmon-supporting thin materials, and in particular, we provide parameters that allow us to readily calculate the response of silver, gold, and graphene islands. Besides their interest for nanoscale electro-optics, the present study emphasizes the great potential of these structures for the design of quantum nanophotonics devices.
\end{abstract}
\maketitle

\section{Introduction}

Plasmons --the collective electron oscillations in nanostructured conductors-- allow us to control light at the nanometer scale, particularly using the large concentration and enhancement of electromagnetic intensity that they generate \cite{LSB03}.  Additionally, and unlike other optical excitations in small systems (e.g., atomic and molecular quantum emitters), plasmons display a powerful combination of two appealing properties: they are robust (i.e., they are not destroyed by the presence of a dielectric environment) and they interact strongly with light (e.g., they display excitation cross-sections typically exceeding the projected area of the nanostructures that sustain the plasmons). These features have facilitated the use of plasmons in applications as varied as nonlinear optics \cite{DN07,DSK08,PN08,MHS14}, ultrasensitive detection down to the single-molecule level via surface-enhanced Raman scattering (SERS) \cite{KWK97,NE97,XBK99,paper125}, cancer diagnosis and therapy \cite{NHH04,LLH05,GLH07,JEE07,JHE08,QPA08}, quantum information processing \cite{CSH06,DSF10,SSR10,paper173}, improved photovoltaics \cite{AP10,WC12}, and subwavelength lithography \cite{DLK14}. Optical metamaterials are also largely relying on subwavelength plasmons to display properties that are not available in naturally occurring materials \cite{E07_2,Z10,BGK10}. These efforts are due in part to the impressive progress made in nanofabrication \cite{NLO09} and colloid chemistry \cite{GPM08,FWB10} techniques, as well as in the theoretical understanding of the response of nanometallic structures \cite{HLC11,paper242}.

The field of plasmonics has been quite focused on noble metals, which are generally regarded as prototypical plasmonic materials, although they suffer from relatively large inelastic losses that limit the lifetime of plasmons down to a few optical cycles in deep-subwavelength structures. In this context, a search for better plasmonic materials has been initiated with a view to reducing absorption \cite{WIN10,FDA10,BA11,NSB13}. Recently, highly doped graphene has emerged as a promising alternative \cite{VE11,paper176,NGG11,JGH11,FAB11,BPV12,paper196,FRA12,paper212,BJS13,paper235}, combining huge field confinement and enhancement with comparatively lower losses \cite{paper212,WLG15}, as well as large electrical tunability of its optical response \cite{WZT08,MSW08,LHJ08,CPB11}. These properties hold great potential for electro-optics applications, such as fast light modulation via electrostatic gating \cite{JGH11,FAB11,paper196,FRA12,paper212,BJS13}, which has been demonstrated with the achievement of frequency variations spanning a whole octave \cite{paper212}.

Unfortunately, plasmons in graphene, as well as in other so-called two-dimensional crystals \cite{SSS13} and in topological insulators \cite{DOL13}, have so far been observed at mid-infrared (mid-IR) and lower frequencies, as they are limited by the low carrier densities in these materials. In contrast, atomically thin metals already possess a substantial conduction electron density in their undoped state, thus sustaining plasmons in the visible and near-infrared (vis-NIR), which are spectral ranges with better prospects for technological applications. Additionally, atomically thin noble metal nanoislands can undergo strong interaction with light and exhibit significant electrical tunability \cite{paper236}, as the doping levels that are currently attainable using gating technology can produce substantial fractional changes in the conduction electron density.

Plasmons in metal clusters of atomic dimensions have been examined and optically characterized for a long time \cite{EP92}, and they have even been used as a toolbox to test the ability of different first-principles computational methods to simulate optical and electron-based spectroscopic measurements \cite{ORR02}.  In a separate effort, atomic self-assembly has been used to produce monoatomic gold wires \cite{LAK01}, which were later shown to sustain extremely confined plasmons \cite{NYI06}. Similar low-dimensional plasmons have been experimentally characterized using electron spectroscopy in ultrathin indium \cite{CKH10} and silicide \cite{RTP10} wires, as well as in few-atomic-layer silver films \cite{MRH99} and monolayer DySi$_2$ \cite{RNP08}. Unfortunately, no further exploration has been pursued towards the coupling of propagating light to these systems and their application to nanophotonics.

Motivated by the availability of these atomically thin materials and their potential for nanophotonics applications, we present here a simple analytical study of the optical properties of disks and ribbons, accompanied by a discussion of their ability to achieve tunable complete optical absorption and quantum strong coupling between plasmons and optical emitters.

\begin{figure}
\begin{center}
\includegraphics[width=130mm,angle=0,clip]{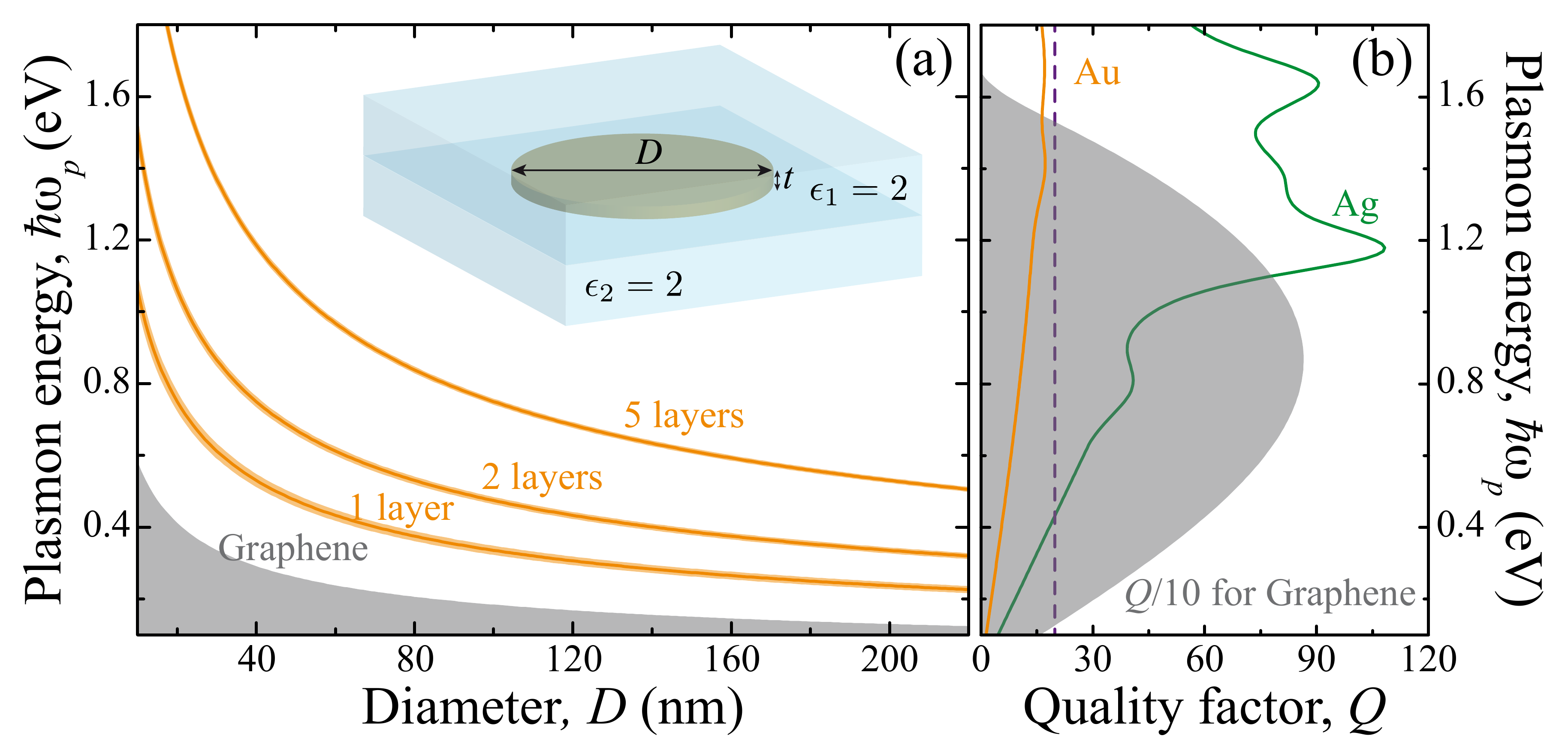}
\caption{{\bf Energy, quality factor, and electrical tunability of thin-disk plasmons.} {\bf (a)} Dipole plasmon energy as a function of diameter for disks formed by 1-5 atomic layers of gold or silver embedded in an $\epsilon=2$ dielectric, as predicted from the analytical model of Table\ \ref{table1}. The thickness of each atomic layer is set to 0.236\,nm (i.e., the separation between (111) atomic planes in these materials, which also have similar values of $\hbar\omegap\approx9\,$eV). Graphene-disk plasmon energies are shown as well for Fermi energies $E_F=0-1\,$eV, illustrating the large degree of electrical tunability of this material. The shaded regions for noble metal disks give the variation of the plasmon energy when electrically doping the disks up to additional carrier densities in the $\pm7\times10^{13}\,$cm$^{-2}$ range (i.e., the same as when doping graphene to $E_F=1\,$eV). {\bf (b)} Quality factor $Q=\omega_p/\gamma$ of gold and silver plasmons in the electrostatic limit. The plasmon damping rate $\gamma$ depends on frequency as shown in Appendix\ \ref{Drudegraph} (see Fig.\ \ref{Fig5}). The quality factor of graphene is obtained from the random-phase approximation conductivity in the local limit (local-RPA \cite{paper235}, see Eq.\ (\ref{localRPA}) in Appendix\ \ref{Drudegraph}), which includes temperature ($T=300\,K$) and interband transition effects. We assume Fermi energies $E_F\le1\,$eV and an intrinsic impurity-limited lifetime estimated for DC mobilities $\mu\le10000\,$cm$^2/(V\,s)$ as $\tau=\mu E_F/ev_F^2$, where $v_F=10^6\,$m/s is the Fermi velocity of graphene. The dashed vertical line indicates the value of $Q$ that equals the full-width fractional variation of the plasmon energy when single-atom gold or silver disks are doped with carrier densities $\pm7\times10^{13}\,$cm$^{-2}$.} \label{Fig1}
\end{center}
\end{figure}

\section{Optical response and tunability of 2D metallic nanoislands}

We describe thin metals in terms of a frequency-dependent 2D conductivity $\sigma(\omega)$, which is related to the bulk dielectric function of the material through $\epsilon(\omega)=1+4\pi\ii\sigma(\omega)/\omega t$, where $t$ is the film thickness. This local approximation works well for atomically thin islands of noble metals with a lateral extension above $\sim10\,$nm, as shown by comparison with quantum-mechanical simulations based upon the random-phase approximation \cite{paper236}. In the low-frequency limit, the dielectric function is well approximated by the Drude model $\epsilon(\omega)=1-\omegapp/\omega(\omega+\ii\gamma)$, where $\omegap$ is the bulk classical plasmon frequency and $\gamma$ is a phenomenological relaxation rate. Combining these two expressions for $\epsilon$, we find the 2D conductivity to reduce to
\begin{equation}
\sigma(\omega)=\frac{\omegapp\,t}{4\pi}\frac{\ii}{\omega+\ii\gamma}.
\label{Drude}
\end{equation}
This formula can even be applied to include the full $\omega$ dependence of the measured dielectric function by simply allowing $\omegap$ and $\gamma$ to depend on $\omega$. In noble metals, these parameters are relatively independent of frequency over the NIR spectral range (see Fig.\ \ref{Fig5} in Appendix\ \ref{Drudegraph}). The present formalism can also describe graphene, where $\omegap$ depends on the Fermi energy $E_F$ relative to the so-called Dirac point as $\omegap=(2e/\hbar)\sqrt{E_F/t}$ (for example, for a nominal graphene thickness $t=0.34\,$nm, as extracted from the interlayer distance in graphite, and considering a realistic value of the Fermi energy $E_F=1\,$eV \cite{CPB11}, we have $\hbar\omegap=4.1\,$eV).

The far-field response of islands that are small compared with the light wavelength can be expressed in terms of their polarizability $\alpha(\omega)$, which admits simple approximate expressions under the reasonable assumption that the lowest-order dipole mode dominates the spectral strength. More precisely, using the expressions derived in Appendix\ \ref{Aana}, we find
\begin{equation}
\alpha(\omega)\approx\frac{tA}{4\pi}\;\frac{\omegapp}{\omega_p^2-\omega(\omega+\ii\gamma)},
\label{alphawp2}
\end{equation}
where $A$ is the area of the island and $\omega_p$ is its lowest-order plasmon frequency (see Table\ \ref{table1} for disks and ribbons). In particular, for a disk of diameter $D$ and thickness $t$ placed at the planar interface between two media of permittivities $\epsilon_1$ and $\epsilon_2$, we have (see Table\ \ref{table1} and derivation in Appendix\ \ref{Aana})
\begin{equation}
\omega_p\approx\frac{\omegap}{\neff}\sqrt{\frac{3\pi t}{8D}},
\label{w1}
\end{equation}
where $\neff=\sqrt{(\epsilon_1+\epsilon_2)/2}$.

Figure\ \ref{Fig1}(a) shows the values of $\omega_p$ predicted by Eq.\ (\ref{w1}) for gold, silver, and graphene disks embedded in silica. We consider noble metal disks consisting of 1, 2, or 5 atomic monolayers, which can clearly reach the NIR. In contrast, the shaded area shows that the plasmon energies  lie in he mid-IR, even for relatively high doping levels ($E_F\le1\,$eV).

We note that the quality factor $Q$ of the plasmon resonances (i.e., $2\pi$ times the number of optical cycles for which the intensity has decayed by $1/\ee$) is given by $\omega_p/\gamma$ and is in fact independent of shape in the electrostatic limit under consideration ($\omega D/c\ll1$). Here, $\gamma$ is the Drude damping of Eq.\ (\ref{Drude}), which depends on frequency and material as shown in Fig.\ \ref{Fig5} (Appendix\ \ref{Drudegraph}), leading to the dependence of $Q$ on $\omega_p$ illustrated by Fig.\ \ref{Fig1}(b). High-quality (mobility $\mu\le10000\,$cm$^2/(V\,s)$) highly doped ($E_F\le1\,$eV) graphene exceeds the performance of gold but is below that of silver for plasmon energies above $\sim1.6\,$eV, which are only reachable with graphene structures that are smaller than those considered in Fig.\ \ref{Fig1} \cite{paper235}, although edge effects can then introduce important corrections \cite{paper183,paper214}.

The range of electro-optical tunability of graphene disk plasmons is illustrated by the shaded area in Fig.\ \ref{Fig1}(a). For a given disk diameter, the plasmon energy can be moved up to the upper value of that area when the doping is increased up to $E_F=1\,$eV. For silver and gold, the range of tunability is lower than in graphene, although it has the advantage that the plasmons are in the NIR. Nonetheless, using currently available gating technology, under the same doping conditions that allow achieving a graphene Fermi energy $E_F=1\,eV$, corresponding to a charge carrier density $n=7\times10^{13}\,$cm$^{-2}$, we obtain a fractional variation of the plasmon energy $\approx\pm n/2n_0=2.5\%$ in gold and silver, where $n_0=1.39\times10^{15}\,$cm$^{-2}$ is the areal density of conduction electrons in these metals. This produces an overall fractional variation of the plasmon that is resolvable with a quality factor $Q\sim n_0/n$, represented by the dashed vertical line of Fig.\ \ref{Fig1}(b), which is clearly within reach with silver.

\begin{figure}
\begin{center}
\includegraphics[width=100mm,angle=0,clip]{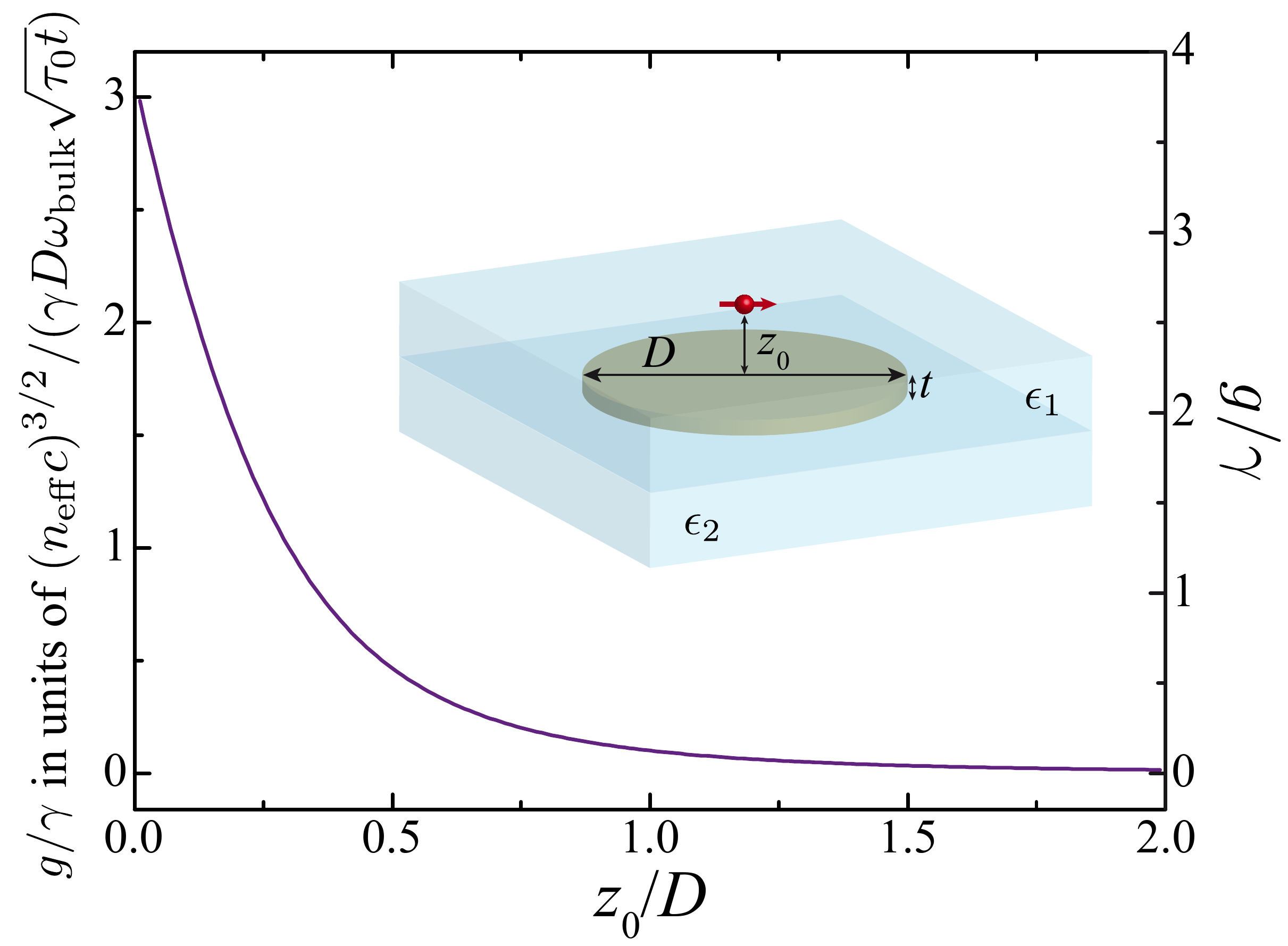}
\caption{{\bf Achieving quantum strong coupling between the lowest-energy plasmon of a thin conducting disk and an optical emitter.} We represent the ratio $g/\gamma$ between the coupling and plasmon decay rates for an emitter oriented parallel to the graphene (see inset), as a function of its separation $z_0$ from the center of a disk of diameter $D$ and thickness $t$. The disk material is characterized by a Drude plasma frequency $\omegap$ and it is placed at the interface between two dielectric media that define the effective permittivity $\neff=\sqrt{(\epsilon_1+\epsilon_2)/2}$. The lifetime of the emitter in the absence of the disk is $\tau_0$. The left vertical axis is given in units of the dimensionless quantity $(\neff c)^{3/2}/[\gamma D \omegap\sqrt{\tau_0 t}]$, whereas the right axis corresponds to the choice
$\hbar\omegap=9\,$eV and $\gamma=0.07\,$eV (gold in the NIR), with $\tau_0=1\,$ns, $D=10\,$nm, $t=0.236\,$nm (i.e., a single  (111) atomic layer), and $\epsilon_1=\epsilon_2=2$ (glass).} \label{Fig2}
\end{center}
\end{figure}

\section{Coupling to Quantum Emitters}

The large concentration of electromagnetic energy associated with the plasmons of atomically thin structures can lead to strong interaction with nearby quantum emitters. This idea has been recently explored in graphene \cite{paper176,paper184,HNG12} and we elaborate on it here to produce a semi-analytical model that is directly applicable to any thin conducting material. For this purpose, we introduce the 2D charge density $\rho_p(\Rb)$ associated with the plasmon as a function of position $\Rb=(x,y)$ along the metal island. This quantity can be conveniently normalized for one plasmon, as discussed in Appendix\ \ref{appendixB}, where analytical expressions are given for the lowest-order dipole modes of disks and ribbons (see Table\ \ref{table1}). Intuitively, $\rho_p$ plays a similar role as the charge density $-e\phi_f^*(\rb)\phi_i(\rb)$ associated with the transition of one electron between the bound states $\phi_i$ and $\phi_f$ of a confined system. We now consider a two-level quantum emitter (e.g., an atom or molecule) of transition dipole $\db_0$. Taking the metal island to lie in the $z=0$ plane and the emitter at position $\rb_0$, the electrostatic emitter-plasmon interaction is simply given by
\begin{equation}
\hbar g=\frac{1}{\neffdos}\int d^2\Rb\,\rho_p(\Rb)\frac{\db_0\cdot(\Rb-\rb_0)}{|\Rb-\rb_0|^3},
\label{hbarg}
\end{equation}
where the integral is extended over the area of the island, $\neff$ is defined right after Eq.\ (\ref{w1}), and $\db_0$ is the effective emitter transition dipole, which is related to its radiative lifetime $\tau_0$ in the absence of the island through $\tau_0^{-1}=4\neff\omega_0^3d_0^2/3\hbar c^3$. Incidentally, $\db_0$ is the transition dipole in vacuum multiplied by a local-field correction $3\neffdos/(2\neffdos+1)$ \cite{YGB1988}.

The quantum evolution of the emitter-plasmon system can be described by the Hamiltonian \cite{paper176,paper184}
\begin{equation}
H=\hbar\left[\omega_pa^+a+\omega_0\sigma^+\sigma
+g(a^+\sigma+a\sigma^+)\right]+\db_p\cdot\Eb^{\rm ext}(t)(a^++a),
\label{HH}
\end{equation}
where $a$ and $\sigma$ ($a^+$ and $\sigma^+$) are the annihilation (creation) operators of the plasmon and the emitter excitation of energies $\hbar\omega_p$ and $\hbar\omega_0$, respectively. Let us stress that we are using the same rate $g$ of emitter-plasmon coupling as defined by the electrostatic energy of Eq.\ (\ref{hbarg}). In the Hamiltonian (\ref{HH}) we are neglecting the direct interaction of the time-dependent external field $\Eb^{\rm ext}$ with the emitter, as its dipole $\db_0$ is assumed to be small compared with the plasmon dipole
\begin{equation}
\db_p=\int d^2\Rb\,\Rb\,\rho_p(\Rb).
\nonumber
\end{equation}
Incidentally, the normalization of $\rho_p$ for a single plasmon is actually based on this dipole, as explained in Appendix\ \ref{appendixB}.

The lifetime of the emitter $\tau_0$ and the plasmon decay rate $\gamma$ can be introduced in the quantum description of the combined system through the density matrix $\rho$, which follows the equation of motion \cite{FT02,MS1990}
\begin{equation}
\frac{d\rho}{dt}=\frac{\ii}{\hbar} \left[\rho,H\right]+\frac{1}{2\tau_0}[2\sigma\rho\sigma^+-\sigma^+\sigma\rho-\rho\sigma^+\sigma]+\frac{\gamma}{2}[2a\rho a^+-a^+a\rho-\rho a^+a].
\label{drhodt}
\end{equation}
In this formalism, we can calculate the polarizability $\alpha(\omega)$ by first obtaining the expected value of the induced dipole from ${\rm tr}\{\db_p(a^++a)\rho\}$ upon illumination with a weak external field $\Eb^{\rm ext}(t)=\Eb_0\ee^{-\ii\omega t}+{\rm c.c.}$ We find the induced dipole to admit the form $\alpha(\omega)\Eb_0\ee^{-\ii\omega t}+{\rm c.c.}$, thus defining $\alpha(\omega)$. In the absence of the emitter (i.e., taking $g=0$), we recover a polarizability $\alpha(\omega)$ as given by the $j=p$ term of Eq.\ (\ref{awdd}), thus demonstrating the self-consistency of our plasmon-normalization scheme. Additionally, when the combined system is considered, the linear polarizability becomes $\alpha(\omega)=\alpha_0(\omega)+\alpha_0^*(-\omega)$ with
\begin{equation}
\alpha_0(\omega)=\frac{d_p^2}{\omega_p-\omega-\ii\gamma/2-g^2(\omega_0-\omega-\ii/2\tau_0)^{-1}}.
\nonumber
\end{equation}
For $\omega_p=\omega_0$, this expression exhibits two poles at frequencies $\omega=\omega_p\pm g$, yielding a vacuum Rabi splitting given by $2g$.

For the vacuum Rabi splitting to be observable, it must be larger than the width of the plasmon peak, that is, $g/\gamma>1$. This condition signals the so-called strong-coupling regime, which has been argued to be achievable in graphene \cite{paper176}. In this regime, the bosonic plasmon state mixes with the fermionic two-level emitter to produce a Jaynes-Cummings ladder of hybridized states \cite{JC1963}, which has been predicted to produce non-classical statistics of the plasmon population upon external illumination, as well as nonlinear optical response \cite{paper184} (i.e., the nonlinearity of the quantum emitter is inherited by the combined plasmon-emitter system).

It should be noted that the decay rate of the excited emitter is enhanced by the coupling to the plasmon and becomes for $\gamma\tau_0\gg1$
\begin{equation}
\Gamma=\frac{1}{\tau_0}+\gamma\,\frac{g^2}{(\omega_p-\omega_0)^2+\gamma^2/4}
\label{GGamma}
\end{equation}
under the condition that the fraction in this expression is small (weak coupling). This well-known result is rederived in Appendix\ \ref{appendixnew} from Eqs.\ (\ref{HH}) and (\ref{drhodt}), and we also show that the dielectric formalism of Appendix\ \ref{Aana} reproduces Eq.\ (\ref{GGamma}) with $g$ as defined by Eq.\ (\ref{hbarg}), provided the plasmon charge density $\rho_p(\Rb)$ is normalized as prescribed in Appendix\ \ref{appendixB}, thus demonstrating the self-consistency of the theoretical methods elaborated in this work.

Equipped with the analytical model for the plasmons of thin conductor islands discussed in the Appendix, we examine in Fig.\ \ref{Fig2} the ratio $g/\gamma$, where $g$ is calculated from Eq.\ (\ref{hbarg}) using the analytical expression of $\rho_p$ for a disk plasmon given in Table\ \ref{table1}. We find $g/\gamma$ to depend on the lifetime of the emitter $\tau_0$, the Drude parameters of the conducting disk $\omegap$ and $\gamma$ (see Eq.\ (\ref{Drude})), the disk diameter and thickness $D$ and $t$, and the index of refraction of the surrounding medium $\neff$ only through a multiplicative coefficient $(\neff c)^{3/2}/(\gamma D \omegap\sqrt{\tau_0 t})$. We plot $g/\gamma$ in Fig.\ \ref{Fig2} expressed in units of that coefficient (left scale) as a function of the distance $z_0$ between the emitter and the center of the disk. The emitter dipole is assumed to be parallel to the graphene. The right scale shows the ratio calculated for $\hbar\omegap=9\,$eV and $\hbar\gamma=0.07\,$eV, typical of gold in the NIR, with $\tau_0=1\,$ns, $D=10\,$nm, $t=0.236\,$nm (i.e., one  (111) atomic layer), and $\neffdos=2$. This result indicates that the strong-coupling regime is reachable over a wide range of distances using gold islands. Silver structures should produce larger coupling because $\gamma$ is smaller in that material (see Fig.\ \ref{Fig5}). Further confinement of the plasmons in structures that display hotspots, such as bowtie antennas \cite{paper216}, could lead to even larger values of $g/\gamma$.

Decoherence produced by inelastic transitions can severely damage the efficiency of quantum emitters when they are placed in a solid-state environment, although close to 100\% efficiencies can be achieved with organic molecules under cryogenic conditions \cite{RWL12}. Fortunately, the enhancement of the coupling rate from the emitter to the plasmon at the frequency of the latter can decrease the relative importance of inelastic decay channels in the emitter (e.g., coupling to phonons of the surrounding material, Auger processes, etc.), so that in practice the coherent part of the decay in emitters such as nitrogen-vacancies in diamond, in which the zero-phonon elastic channel accounts for only a small fraction of the emission, can be enhanced by coupling to the nanoisland, and we are thus under similar conditions as those considered in this study (i.e., the emitter decay through coupling to a plasmon dominates over other inelastic channels).

\section{Complete Optical Absorption}

Complete optical absorption has been studied and observed over many frequency ranges in disordered metal films \cite{HM1973,KNG10}, through lattice resonances in gratings and planar metamaterials \cite{HM1976,MP1976,PT1990,TSP00,GCJ02,LSM08,paper090}, assisted by localized plasmonic resonances \cite{KSP06,paper107}, using multilayer structures \cite{CPT04}, and in overdense plasma \cite{BFS05}. However, the possibility of achieving complete optical absorption in atomically thin films offers additional advantages, as we discuss below.

It is well known that the maximum absorbance produced by an optically thin film in a homogeneous environment is 50\%: the incident light induces charges and currents that have no memory of where light is coming from, and therefore, they radiate symmetrically towards both sides of the film with a scattered wave amplitude $r$; consequently, the reflected and transmitted amplitudes are $r$ and $1+r$, where the first term in the transmission is the incident field of unit amplitude; the absorbance is thus $1-|r|^2-|1+r|^2$, whose maximum value is $1/2$ as a function of the complex variable $r$. Now, with the addition of a reflecting screen on one side of the film, light can make two passes through the thin material, producing a maximum of 100\% absorption if both incident and reflected waves are in phase at that plane. This is the so-called Salisbury screen configuration, in which the film/metal-screen separation should be roughly $\lambda/4n$ \cite{FM1988,E02} (i.e., a phase of $\pi$ is produced by the metallic reflection and another $\pi$ contribution comes from phase associated with the round trip propagation between the film and the screen, assumed to be embedded in an environment of refractive index $n$).

These ideas have been recently explored for graphene, leading to the prediction of complete optical absorption by a suitably patterned carbon layer \cite{paper182,FP12}, under the condition that the extinction cross-section per unit cell element is of the order of the unit cell area. The observation of electrically tunable large absorbance in patterned graphene has been recently accomplished \cite{paper230,JBS14}. We argue here that similar levels of tunable absorption are achievable using noble metals.

It is instructive to first examine the maximum extinction of a thin island in vacuum. The corresponding cross section is \cite{V1981} $\sigma^{\rm ext}=4\pi(\omega/c){\rm Im}\{\alpha(\omega)\}$, which upon insertion of Eq.\ (\ref{alphawp2}) is found to exhibit a maximum at $\omega=\omega_p$, given approximately by $\sigma^{\rm ext}=(\omegapp t/\gamma c)\times A$. Remarkably, this maximum extinction is independent of shape for a given area $A$ of the island, under the assumption that an individual plasmon mode dominates the extinction. In particular, for single atomic layers of gold (silver) films ($t=0.236\,$nm), considering the plasmon energy $\hbar\omega_p$ to be in the NIR, we have $\hbar\omegap\sim9\,$eV and $\hbar\gamma\sim70\,$meV ($\hbar\gamma\sim20\,$meV), so that the maximum cross section is $1.4$ ($4.8$) times the area of the island. For highly doped graphene, this number is even larger due to the comparatively lower losses of this material (see Fig.\ \ref{Fig5}).

\begin{figure}
\begin{center}
\includegraphics[width=110mm,angle=0,clip]{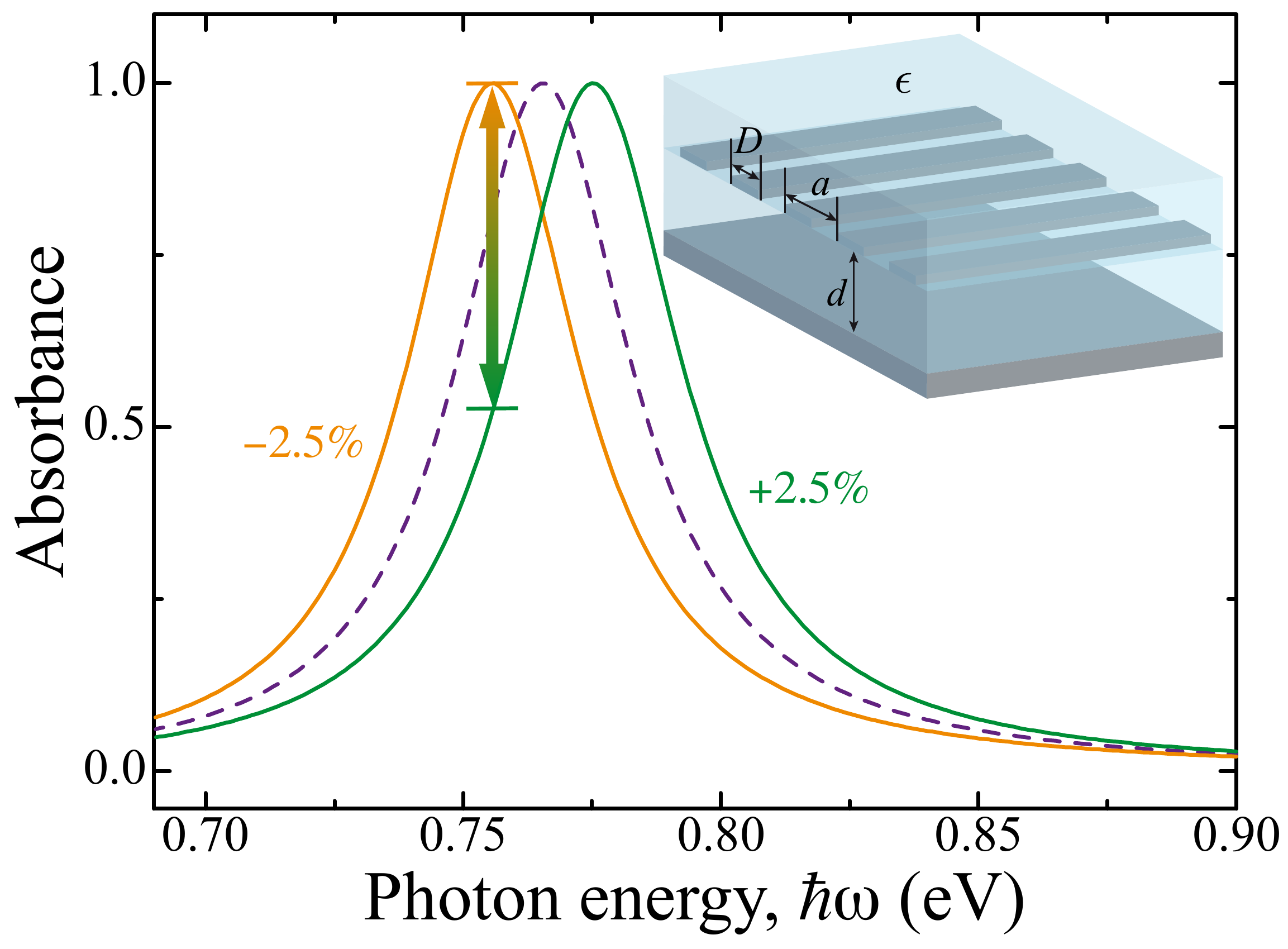}
\caption{{\bf Tunable complete optical absorption.} We show the absorbance of a single-atomic-layer ribbon array embedded in silica ($\epsilon=n^2=2$) and placed at a distance $d=286\,$nm above a perfectly reflecting mirror. The ribbon width and period are $D=20\,$nm and $a=68\,$nm, respectively. The response of silver is described with $\hbar\omegap=9\,$eV and $\hbar\gamma=0.02\,$eV. The dashed curve shows the result for undoped silver ribbons, whereas the solid curves correspond to a fractional variation of $\pm2.5\%$ in the areal conduction electron density (i.e., $\pm7\times10^{13}\,$cm$^{-2}$), enabling a large change in absorption, as indicated by the double vertical arrow.} \label{Fig3}
\end{center}
\end{figure}

Complete optical absorption is achievable in periodic arrays placed above a Salisbury screen. The normal-incidence reflection coefficient of a doubly-periodic array of small period $a$ compared with the light wavelength, surrounded by a homogeneous environment of refractive index $n$, reduces to \cite{paper182}
\begin{align}
r=\frac{\ii S}{\alpha^{-1}-G},
\nonumber
\end{align}
where $S=(2\pi\omega/nA_cc)$, $G=\mathfrak{g}/n^2a^3+\ii S$, $A_c$ is the unit cell area, and $\mathfrak{g}$ is a number that depends on symmetry (e.g., $\mathfrak{g}\approx5.52$ and $\mathfrak{g}\approx4.52$ for hexagonal and square arrays, respectively, assuming that all islands interact through their induced dipoles; corrections due to nearest-neighbor interactions beyond dipolar terms are possible for closely spaced islands, in which case the coefficient $\mathfrak{g}$ can depend on their shape). With a Salisbury screen of reflectivity $r_0=|r_0|\ee^{\ii\varphi_0}$ separated a distance $d$ from the array, the incident and reflected waves are exactly on phase when $\varphi_0+2\omega dn/c$ is a multiple of $2\pi$. Complete absorption is then produced under the condition $|r_0|=-r/(1+2r)$, which is satisfied at a frequency $\omega$ given by $\omega^2=\omega_p^2-(\mathfrak{g}/4\pi n^2)(tA/a^3)\omegapp$, where $\omega_p\approx(\omegap/n)\sqrt{3\pi t/8D}$, provided we have
\begin{align}
\frac{2n\gamma c}{\omegapp t}\,\frac{A_c}{A}=1+\frac{1}{|r_0|}.
\label{cond}
\end{align}
Interestingly, this condition for complete optical absorption is also independent of the shape of the island. Using the approximate values of $\omegap$ and $\gamma$ noted above for gold and silver in the NIR, and considering for simplicity a non-absorbing Salisbury screen ($|r_0|=1$) and a glass environment ($n^2=2$), the condition (\ref{cond}) for perfect absorption in gold (silver) arrays is fulfilled with a fraction $A/A_c=1.02$ (0.29) between the areas of the island and the unit cell. Consequently, this condition can be easily met using silver atomic monolayers, and also with multilayers of either gold or silver.

Incidentally, similar results are obtained for ribbon arrays of period $a$ \cite{paper235}. Then, the condition (\ref{cond}) remains unchanged, with $A/A_c=D/a$, where $D$ is the ribbon width. Because of the ribbon translational symmetry, we work with 2D rather than 3D scattering, so we need to redefine $\alpha(\omega)\approx(tD/4\pi)\,\omegapp/[\omega_p^2-\omega(\omega+\ii\gamma)]$, $G=2\pi^2/3n^2 a^2+\ii S$ and $S=2\pi\omega/nac$. The frequency at which complete absorption occurs is then $\omega=\omega_p\sqrt{1-(\pi^2/24)(D/a)^2}$, where $\omega_p\approx(\omegap/n)\sqrt{4t/\pi D}$ (see Table\ \ref{table1}).

It is important to note that the value of $\omegap$ can be modulated by $\approx\pm2.5\%$ using currently available gating technology (see above), and this in particular produces peak shifts larger than the peak width in silver (see Fig.\ \ref{Fig1}(b)), as shown in Fig.\ \ref{Fig3} for an illustrative example. This type of structure is convenient because the ribbons can be contacted at a large distance away from the region in which the optical modulation is pursued. Besides, the strong absorption only occurs for polarization across the ribbons, thus suggesting a possible application as tunable polarizers.

\section{Outlook and Perspectives}

Graphene plasmons are focusing much attention due in part to the demonstrated ability to modulate their frequencies by electrically doping the carbon layer through suitably engineered gates \cite{JGH11,FAB11,paper196,FRA12,paper212,BJS13}. This modulation can be potentially realized at high speeds, as the number of charge carriers that are needed to produce changes in the Fermi energy of the order of the electronvolt is relatively small, and consequently, so they are the inductance and capacitance associated with the graphene itself. Unfortunately, graphene plasmons have only been observed at mid-infrared and lower frequencies. Moving to the vis-NIR is challenging and requires patterning structures with sizes $<10\,$nm under realistically attainable doping conditions. In this respect, molecular self-assembly provides a viable way of synthesizing nanographenes in this size range \cite{LWZ08,LTA13,CRJ10,M14}. Doped carbon nanotubes have also been predicted to display plasmons that are rather insensitive to their degree of chirality \cite{paper235}, and therefore, they provide a viable route towards the fabrication of large-scale tunable plasmonic structures operating in the vis-NIR regime. Polycyclic aromatic hydrocarbons also sustain excitations at vis-NIR frequencies that behave as graphene plasmons \cite{paper215} and constitute promising candidates to advance towards atomic-scale tunable plasmonics.

Although plasmons in atomically thin metals have been observed in several systems \cite{NYI06,CKH10,RTP10,MRH99,RNP08}, these studies have focused on extended surfaces whose plasmon dispersion relations are far from the light cone, thus averting the possibility of direct coupling to propagating light. Further patterning of these types of surfaces into disks and ribbons such as those considered here could facilitate the coupling to optical probes. An alternative option consists in decorating atomically thin films with dielectric colloids to provide periodic optical contrast. Substrate pre-patterning of disks, ribbons, or other morphologies, followed by atomic layer deposition constitutes yet another possibility.

We conclude that atomically thin materials hold great potential for the manipulation of light at truly nanometer scales and for the development of applications to optical signal processing, quantum optics, and sensing. We should emphasize that small nanoparticles, not necessarily atomically thin, can produce similar levels of strong-coupling and optical absorption as discussed above for thin films, although they are not tunable using electrical gates because the injected charge carriers have to compete with a much larger number of bulk conduction electrons. The great opportunities offered by these materials are however accompanied by formidable challenges to produce the islands at designated positions and with controlled morphology, possibly requiring a combination of top-down patterning and bottom-up self-assembly methods similar to those mentioned above.

\appendix

\section{OPTICAL RESPONSE OF THIN METAL ISLANDS IN THE ELECTROSTATIC LIMIT}
\label{Aana}

The scale-invariant character of the electrostatic problem (i.e., the absence of a frequency-dependent length scale imposed by the wavelength) has been used on several ocations to express the solutions in terms of modal expansions \cite{B1979,GF1988,OI1989,BD92}. Here, we formulate a suitable decomposition for optically thin structures. We consider islands of small characteristic size $D$ (e.g., the diameter for disks or the width for ribbons) compared with the light wavelength, such that the optical electric field $\Eb=-\nabla\phi$ can be expressed in terms of a scalar potential $\phi$. The islands are however taken to be large enough to be described as infinitesimally thin domains characterized by a local, frequency-dependent 2D conductivity $\sigma(\omega)$. Following previous analyses for graphene \cite{paper212,paper228,paper235}, we write the self-consistent potential at positions $\Rb=(x,y)$ in the plane of the island as
\begin{equation}
\phi(\Rb)=\phi^{\rm ext}(\Rb)+\frac{1}{\neffdos} \frac{\ii}{\omega} \int \frac{d^2\Rb'}{|\Rb-\Rb'|}\,\nabla_{\Rb'}\cdot\sigma(\Rb',\omega)\nabla_{\Rb'}\phi(\Rb'),
\label{eq1}
\end{equation}
which is the sum of the external perturbation $\phi^{\rm ext}$ and the contribution produced by the induced charges (integral term). The island is chosen to lie at the planar interface between two media of permittivities $\epsilon_1$ and $\epsilon_2$, which contribute to the above expression through a $1/\neffdos$ factor multiplying the in-plane Coulomb potential, where
\begin{equation}
\neff=\sqrt{(\epsilon_1+\epsilon_2)/2}.
\label{nefflabel}
\end{equation}
Now, using the definitions $\th=\Rb/D$ and \[\eta=\frac{1}{\neffdos}\frac{\ii\sigma(\omega)}{\omega D},\] taking the gradient in both sides of Eq.\ (\ref{eq1}), and multiplying by $-\sqrt{f}$, we find \cite{paper228}
\begin{equation}
\fE(\th,\omega)=\fE^{\rm ext}(\th,\omega)+\eta(\omega)
\int d^2\th'\;\Mb(\th,\th')\cdot\fE(\th',\omega),
\label{fEfE}
\end{equation}
where \[\fE(\th,\omega)=-\sqrt{f(\th)}\,{\nabla_{\th}}\phi(\th,\omega),\] whereas $f(\th)$ is a filling function that is 1 if $\th$ lies on the metal and zero otherwise, so that the frequency and spatial dependences of the conductivity are separated as $\sigma(\Rb,\omega)=f(\Rb)\sigma(\omega)$. Notice that this formalism is also valid for inhomogeneous layers by allowing $f$ to take values different from 0 or 1 \cite{paper194}. Here, we have defined $\Mb(\th,\th')=\sqrt{f(\th)f(\th')}\;\;{\nabla_{\th}}\otimes{\nabla_{\th}}\,(1/|\th-\th'|)$, which is a real, symmetric operator that admits a complete orthonormal set of real eigenvectors $\fE_j$ and eigenvalues $1/\eta_j$ satisfying the relations
\begin{align}
{\rm eigensystem} &\;\;\;\rightarrow\;\;\; \eta_j\int d^2\th'\;\Mb(\th,\th')\cdot\fE_j(\th')=\fE_j(\th), \nonumber\\
{\rm orthogonality} &\;\;\;\rightarrow\;\;\; \int d^2\th\;\fE_j(\th)\cdot\fE_{j'}(\th)=\delta_{jj'}, \nonumber\\
{\rm closure} &\;\;\;\rightarrow\;\;\; \sum_j\fE_j(\th)\cdot\fE_{j}(\th')=\delta(\th-\th')\mathcal{I}_2. \nonumber
\end{align}
Then, the solution to Eq.\ (\ref{fEfE}) reduces to
\begin{equation}
\fE=\sum_j[c_j/(1-\eta/\eta_j)]\fE_j,
\label{fEsolution}
\end{equation}
where $c_j=\int d^2\th\;\fE^{\rm ext}(\th)\cdot\fE_{j}(\th)$.

Applying these results to a uniform external field $\Eb_0$ aligned with a symmetry direction of the island $\xx$ (i.e., for $\fE^{\rm ext}=\sqrt{f}D\,E_0\xx$), we obtain the polarizability along that direction $\alpha(\omega)=E_0^{-1}\int d^2\Rb\,x\rho^{\rm ind}(\Rb)$ from the induced density
\begin{equation}
\rho^{\rm ind}(\Rb)=\frac{\ii\sigma}{\omega}\nabla_\Rb\cdot f(\Rb) \nabla_\Rb\phi(\Rb)=\frac{-\ii\sigma}{\omega D^2}\nabla_{\th}\cdot\sqrt{f(\th)} \fE(\th).
\label{rhoind}
\end{equation}Inserting Eq.\ (\ref{fEsolution}) into this expression, we obtain
\begin{equation}
\alpha(\omega)=D^3\sum_j\frac{A_j}{\frac{-1}{\neffdos\eta_j}-\frac{\ii\omega D}{\sigma(\omega)}},
\label{alpha}
\end{equation}
where $j$ runs over eigenmodes of the system and
\begin{equation}
A_j=\left|\int d^2\th\;\sqrt{f(\th)}\;\mathcal{E}_{jx}(\th)\right|^2
\label{Aj}
\end{equation}
are dimensionless coupling coefficients. Using the conductivity of Eq.\ (\ref{Drude}), we can recast Eq.\ (\ref{alpha}) as
\begin{equation}
\alpha(\omega)=\frac{tD^2}{4\pi}\sum_j\frac{A_j\omegapp}{\omega_j^2-\omega(\omega+\ii\gamma)},
\label{alphabis}
\end{equation}
where the plasmons of the nanoisland can be identified with modes $j$ of negative eigenvalues $\eta_j$ and frequencies
\begin{equation}
\omega_j=\frac{\omegap}{\neff}\;\frac{1}{\sqrt{-4\pi\eta_j}}\;\sqrt{\frac{t}{D}},
\label{wj}
\end{equation}
corresponding to the $\omega\approx\omega_j-\ii\gamma/2$ poles of Eq.\ (\ref{alphabis}).

\begin{table*}
\begin{center}
\begin{tabular}{lcccc}
\hline \\               &\;\; $\omega_p$  \;\;&\;\; $\eta_p$ \;\;&\;\; $A_p$ \;\;\;\;&\;\;\;\; $\rho_p(\Rb)$ \\ \\ \hline \\
disk
&\;\; $(\omegap/\neff)\sqrt{3\pi t/8D}$                                    \;\;
&\;\; $-2/3\pi^2$ \;\;
&\;\; $\pi/4$ \;\;\;\; &\;\;\;\; $C_{\rm d}\frac{x}{D}\,\frac{1+(1/4)\ee^{-5(1-2R/D)}}{\sqrt{1-2R/D}}$ \\ \\ \hline \\
ribbon
&\;\; $(\omegap/\neff)\sqrt{4t/\pi D}$                                    \;\;
&\;\; $-1/16$ \;\;
&\;\; $L/D$ \;\;\;\; &\;\;\;\; $C_{\rm r}\frac{x}{D}\,\frac{1+(1/4)\ee^{-5[1-(2x/D)^2]}}{\sqrt{1-(2x/D)^2}}$   \\ \\ \hline
\end{tabular}
\end{center}
\caption{Analytical approximations for the parameters $\omega_p$, $\eta_p$, and $A_p$ corresponding to the lowest-order dipole resonance of a disk of diameter $D$ and a ribbon of width $D$ and length $L\rightarrow\infty$. The dependence on dielectric environment is entirely contained in a factor $\neff=\sqrt{(\epsilon_1+\epsilon_2)/2}$ dividing the frequency when the film is placed at the interface between two media of permittivities $\epsilon_1$ and $\epsilon_2$. The rightmost column shows approximations to the radial and transversal dependences of the charge density induced by a single plasmon in disks and ribbons, respectively, which involve the coefficients $C_{\rm d}^2=4.58\,\neff\,(\hbar\omegap/D^3)\sqrt{t/D}$ and $C_{\rm r}^2=3.10\,\neff \,(\hbar\omegap/LD^2)\sqrt{t/D}$. The ribbon edges are taken at $x=\pm D/2$.}
\label{table1}
\end{table*}

Equation\ (\ref{alphabis}) involves coefficients that are subjected to two useful sum rules \cite{paper235}:
\begin{enumerate}[{\bf i.}]
\item For any arbitrarily shaped island of area $A$, we have $\sum_j A_j=A/D^2$. This result is readily obtained from the definition of the $A_j$ coefficients in Eq.\ (\ref{Aj}) upon application of the closure relation for $\fE_j$ (see above). Applying this sum to a Drude metal (i.e., for frequency-independent $\omegap$), we conclude that the integral of the extinction cross-section ($\propto\omega\alpha(\omega)$) is actually proportional to $\omegap A t$, which is in turn proportional to the number of electrons (i.e., it fulfills the $f$-sum rule \cite{PN1966}).
\item Another sum rule follows from Eq.\ (\ref{alpha}) in the $\omega\rightarrow0$ limit (i.e., when the island behaves as a perfect conductor, so that $\omega/\sigma\rightarrow0$). Without loss of generality, we can consider a freestanding island ($\neff=1$), so we have $-\sum_j\eta_jA_j=\alpha(0)/D^3$. Now, for in-plane polarization of a disk of diameter $D$, we have $\alpha(0)=D^3/6\pi$ (this result can be derived from the polarizability of an ellipsoid of vanishing height \cite{paper090}), which leads to $-\sum_j\eta_jA_j=1/6\pi$. Likewise, from the transversal polarizability of a thin metal ribbbon of width $D$ \cite{V1981} ($\alpha(0)=D^2L/16$, where $L\rightarrow\infty$ is the length), we find $-\sum_j\eta_jA_j=L/16D$.
\end{enumerate}
Interestingly, for these types of structures and polarizations, we find one single mode $j=p$ to be dominant and to absorb most of the weight in the above sums \cite{paper235}. More precisely, this is the lowest-order dipole plasmon. Neglecting all other modes, these sum rules lead to the values of $\omega_p$, $\eta_p$, and $A_p$ listed in Table\ \ref{table1} and extensively used throughout this work to produce analytical estimates of plasmonic behavior.

\begin{figure}
\begin{center}
\includegraphics[width=80mm,angle=0,clip]{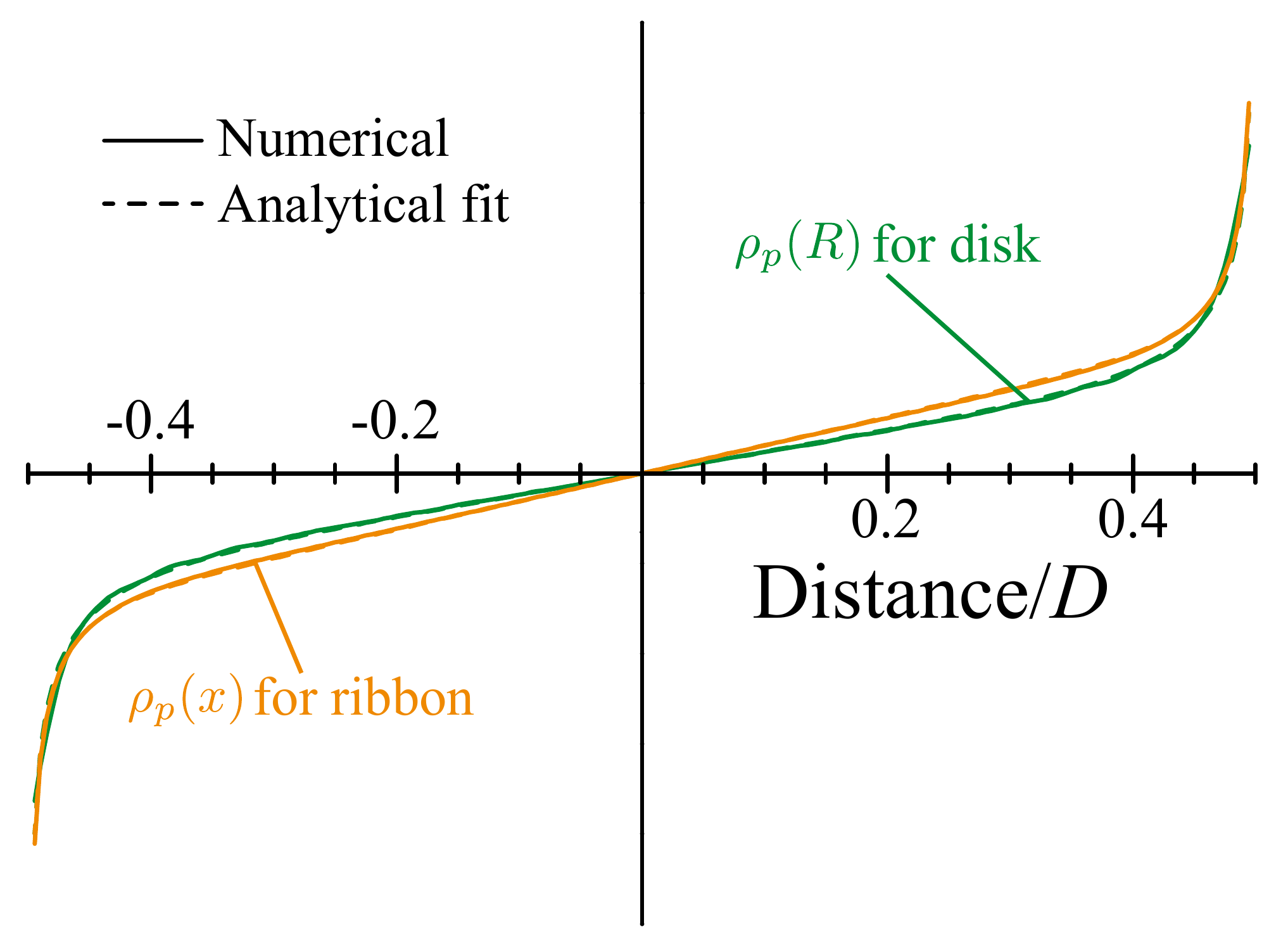}
\caption{{\bf Induced charge of a single plasmon in thin disks and ribbons.} We show the charge densities associated with the lowest-order dipolar plasmon in a disk of diameter $D$ ($m=1$ azimuthal symmetry) and in a ribbon of width $D$ (transversal polarization with wave vector $k_\parallel=0$ along the direction of translational symmetry). The solid curves correspond numerical results taken from the literature for disks \cite{paper237} and ribbons \cite{paper194}. The dashed curves represent the fitting functions of Table\ \ref{table1}.} \label{Fig4}
\end{center}
\end{figure}

\section{CHARGE INDUCED BY A SINGLE PLASMON}
\label{appendixB}

We introduce a purely electrostatic scheme to normalize the induced charge density $\rho_j(\Rb)$ associated with a single plasmon $j$. From linear-response theory \cite{PN1966}, the polarizability reads
\begin{equation}
\alpha(\omega)=\frac{1}{\hbar}\sum_j\,d_j^2\,\left(\frac{1}{\omega_j-\omega-\ii\gamma/2}+\frac{1}{\omega_j+\omega+\ii\gamma/2}\right),
\label{awdd}
\end{equation}
where
\begin{equation}
d_j=\int d^2\Rb\;x\;\rho_j(\Rb)
\label{dj}
\end{equation}
is the dipole moment associated with mode $j$ for polarization along a symmetry direction $\xx$. Now, in order to compare Eq.\ (\ref{awdd}) with Eq.\ (\ref{alphabis}), we neglect $\gamma^2$ in front of $\omega_j^2$ and approximate Eq.\ (\ref{awdd}) as
\begin{equation}
\alpha(\omega)\approx\frac{2}{\hbar}\sum_j\frac{\omega_jd_j^2}{\omega_j^2-\omega(\omega+\ii\gamma)}.
\label{bbb}
\end{equation}
Now, inserting the Drude conductivity of Eq.\ (\ref{Drude}) into Eq.\ (\ref{alpha}), comparing the result with Eq.\ (\ref{bbb}), and taking Eq.\ (\ref{wj}) into account, we find the normalization condition
\begin{equation}
d_j^2=A_j\hbar\omegap\;\neff\;\sqrt{\left(\frac{-\eta_j}{16\pi}\right)\;tD^5}.
\label{dj2}
\end{equation}
Within the single-mode approximation noted at the end of the previous paragraph, writing the charge density associated with the lowest-order disk dipole plasmon as $\rho_p(\Rb)=\rho_p(R)\cos\varphi$, where $\rho_p(R)$ gives the radial dependence, we find the normalization condition
\begin{equation}
\left|\int_0^{D/2} R^2dR\;\rho_p(R)\right|^2=\frac{\hbar\omegap\neff}{16\pi^2}\;\sqrt{\frac{2tD^5}{3\pi}}.
\nonumber
\end{equation}
Similarly, the plasmon charge density $\rho_p(x)$ along the transversal direction $\xx$ of a ribbon contained in the $|x|<D/2$ region satisfies
\begin{equation}
\left|\int_0^{D/2} x\,dx\;\rho_p(x)\right|^2=\frac{\hbar\omegap\neff}{64A}\;\sqrt{\frac{tD^5}{\pi}},
\nonumber
\end{equation}
where $A=LD$ is the ribbon area (with $L\rightarrow\infty$) and we have utilized the symmetry $\rho_p(-x)=-\rho_p(x)$. Using these normalizations, we find that the analytical expressions for $\rho_p$ that are given in Table\ \ref{table1}, where the density profile is taken to fit previous calculations \cite{paper237,paper194} based upon the boundary-element method. Actually, these formulas reproduce rather well the calculated density profiles, as shown in Fig.\ \ref{Fig4}.

An alternative normalization is provided by the fact that the plasmon energy $\hbar\omega_j$ is twice its electrostatic energy. This condition can be expressed as
\begin{equation}
\hbar\omega_j=\frac{2}{\neffdos}\int d^2\Rb\int d^2\Rb' \frac{\rho_j(\Rb)\rho_j(\Rb')}{|\Rb-\Rb'|},
\nonumber
\end{equation}
which leads to values of the normalization coefficients $C_{\rm d}^2=4.48\,(\hbar\omegap/D^3) \sqrt{t/D}$ and $C_{\rm r}^2=3.00\,(\hbar\omegap/LD^2)\neff\sqrt{t/D}$ for disks and ribbons, respectively, in excellent agreement with those shown in the caption of Table\ \ref{table1}, considering that we are making the approximation that only the lowest-order plasmon contributes to the response. Additionally, we show in Appendix\ \ref{appendixnew} that the plasmon normalization here introduced is the same as that needed to describe the coupling rate between the plasmon and an optical emitter through the intuitive expression given in Eq.\ (\ref{hbarg}).

\section{PLASMON-ENHANCED EMITTER DECAY RATE}
\label{appendixnew}

\subsection{Density-matrix approach}

It is convenient to expand the density matrix of Eq.\ (\ref{drhodt}) as
\begin{equation}
\rho=\sum_{ln,l'n'}\rho_{ln,l'n'}(t)\;\ee^{-\ii(l-l')\omega_0t}\;\ee^{-\ii(n-n')\omega_pt}\;\ee^{-(1/2\tau_0)(l+l')t}\;\ee^{-(\gamma/2)(n+n')t}\;|ln\rangle\langle l'n'|,
\label{rhoexp}
\end{equation}
where $\rho_{ln,l'n'}(t)$ are time-dependent coefficients, while $|ln\rangle$ denotes a state with $n$ plasmons accompanied by the excited (de-excited) emitter for $l=1$ ($l=0$). Inserting this expression into Eq.\ (\ref{drhodt}), we find
\begin{align}
\dot{\rho}_{ln,l'n'}=-\ii g\bigg[&\rho_{0n+1,l'n'}\sqrt{n+1}\;\ee^{-\ii\Delta t}\delta_{l,1}
+\rho_{1n-1,l'n'}\sqrt{n}\;\ee^{\ii\Delta t}\delta_{l,0} \nonumber\\
-&\rho_{ln,0n'+1}\sqrt{n'+1}\;\ee^{\ii\Delta^* t}\delta_{l',1}
-\rho_{ln,1n'-1}\sqrt{n'}\;\ee^{-\ii\Delta^* t}\delta_{l',0}\bigg] \nonumber\\
+\;\tau_0^{-1}\;&\rho_{1n,1n'}\;\ee^{-t/\tau_0}\delta_{l,0}\delta_{l',0}+\;\gamma\;\rho_{ln+1,l'n'+1}\sqrt{(n+1)(n'+1)}\;\ee^{-\gamma t},
\label{eqmot}
\end{align}
where $\Delta=\omega_p-\omega_0-\ii(\gamma-\tau_0^{-1})/2$.
We now argue that, for an initial density matrix in which all terms of Eq.\ (\ref{rhoexp}) with $l+n>N$ or $l'+n'>N$ are zero (i.e., a density matrix involving a maximum number of $N$ excitations in the combined plasmon-emitter system), the last term of Eq.\ (\ref{eqmot}) vanishes because it involves states that are never populated. We are then left with a self-contained subset of equations involving coefficients $\rho_{ln,l'n'}$ with $l+n=l'+n'=N$. For this manifold of $N$ excitations, we trivially find solutions $\rho_{ln,l'n'}=a_{ln}a^*_{l'n'}$, where the coefficients $a_{ln}$ satisfy the equations
\begin{align}
&\dot{a}_{0N}=-\ii g\,a_{1N-1}\sqrt{N}\;\ee^{\ii\Delta t}\nonumber\\
&\dot{a}_{1N-1}=-\ii g\,a_{0N}\sqrt{N}\;\ee^{-\ii\Delta t}
\nonumber
\end{align}
and admit the familiar Jaynes-Cummings solutions \cite{JC1963}
\begin{equation}
\left(\begin{array}{cc}
a_{0N} \\
a_{1N-1}
\end{array} \right)
\propto\left(\begin{array}{cc}
-\omega_\pm\,\ee^{-\ii\omega_\mp t}\\
g\sqrt{N}\,\ee^{\ii\omega_\pm t}
\end{array} \right)
\nonumber
\end{equation}
with $\omega_\pm=-\Delta/2\pm\sqrt{\Delta^2/4+g^2N}$. In the $|\Delta|\gg g$ limit, we have $\omega_+\approx g^2N/\Delta$ and $\omega_-\approx-\Delta$, so the solution with the upper (lower) signs has $|a_{0N}|\ll|a_{1N-1}|$ ($|a_{0N}|\gg|a_{1N-1}|$) at $t=0$, and therefore it corresponds to the initially excited (de-excited) emitter. The decay rate of the emitter when it is initially excited and the plasmon is not populated (i.e., starting from $l=1,n=0$) is then given by the decay of the $|10\rangle\langle10|$ term of Eq.\ (\ref{rhoexp}) in the upper-sign solution. We find $\Gamma=\tau_0^{-1}-\dot{\rho}_{10,10}(0)/\rho_{10,10}(0)=\tau_0^{-1}+2\,{\rm Im}\{\omega_+\}\approx\tau_0^{-1}+2g^2\,{\rm Im}\{1/\Delta\}$, which reduces to Eq.\ (\ref{GGamma}) under the condition $\gamma\tau_0\gg1$.

\subsection{Dielectric approach}

The decay rate of an emitter placed at a position $\rb_0$ in the vicinity of the plasmonic structure can be related to its transition dipole $\db_0$ as \cite{NH06}
\begin{equation}
\Gamma=\frac{1}{\tau_0}+\frac{2}{\hbar}\,{\rm Im}\{\db_0^*\cdot\Eb^{\rm ind}\},
\label{DA1}
\end{equation}
where $\Eb^{\rm ind}$ is the self-induced electric field produced by a dipole $\db_0$ located at $\rb_0$. We can calculate this field from the dielectric formalism of Appendix\ \ref{Aana} using the induced density of Eq.\ (\ref{rhoind}), but now the coefficients $c_j$ of Eq.\ (\ref{fEsolution}) have to be obtained from the external dipole field $\Eb^{\rm ext}=(1/\neffdos)(\db_0\cdot\nabla_0)\nabla_0(1/|\rb_0-\Rb|)$, where $\neff$ is defined in Eq.\ (\ref{nefflabel}). (Notice that the potential produced by the dipole at the planar interface between media of permittivities $\epsilon_1$ and $\epsilon_2$ is the same as in vacuum multiplied by $1/\neffdos$.) After some algebra, we find
\begin{equation}
\db_0^*\cdot\Eb^{\rm ind}=\frac{t}{4\pi n_{\rm eff}^4D^4}\sum_j\frac{\omegapp}{\omega_j^2-\omega(\omega+\ii\gamma)}
\left|\int d^2\Rb\;\left[\nabla_{\th}\cdot\sqrt{f(\th)}\fE_j(\th)\right]\left(\db_0\cdot\nabla_0\frac{1}{|\rb_0-\Rb|}\right)\right|^2.
\label{DA2}
\end{equation}
Inserting Eq.\ (\ref{DA2}) into Eq.\ (\ref{DA1}), and retaining only the $j=p$ term, we recover the emitter decay rate given by Eq.\ (\ref{GGamma}) with the exact same definition of the coupling rate $g$ as given by Eq.\ (\ref{hbarg}), provided we define
\begin{equation}
\rho_j(\th)=\sqrt{\frac{\hbar\omegapp t}{8\pi\omega_jD^4}}\;\left[\nabla_{\th}\cdot\sqrt{f(\th)}\fE_j(\th)\right].
\nonumber
\end{equation}
Finally, inserting this expression into Eq.\ (\ref{dj}), integrating by parts, and keeping in mind the definition of $A_j$ in Eq.\ (\ref{Aj}), we recover Eq.\ (\ref{dj2}) for the plasmon transition strength. Therefore, we conclude that the normalization of the plasmon charge density discussed in Appendix\ \ref{appendixB}, based upon the polarizability of the plasmonic structure, produces the same decay rate of a neighboring emitter when calculated either following the semi-classical dielectric formalism described in this paragraph or using the density-matrix formalism with the intuitive coupling rate defined by Eq.\ (\ref{hbarg}).

\begin{figure}
\begin{center}
\includegraphics[width=100mm,angle=0,clip]{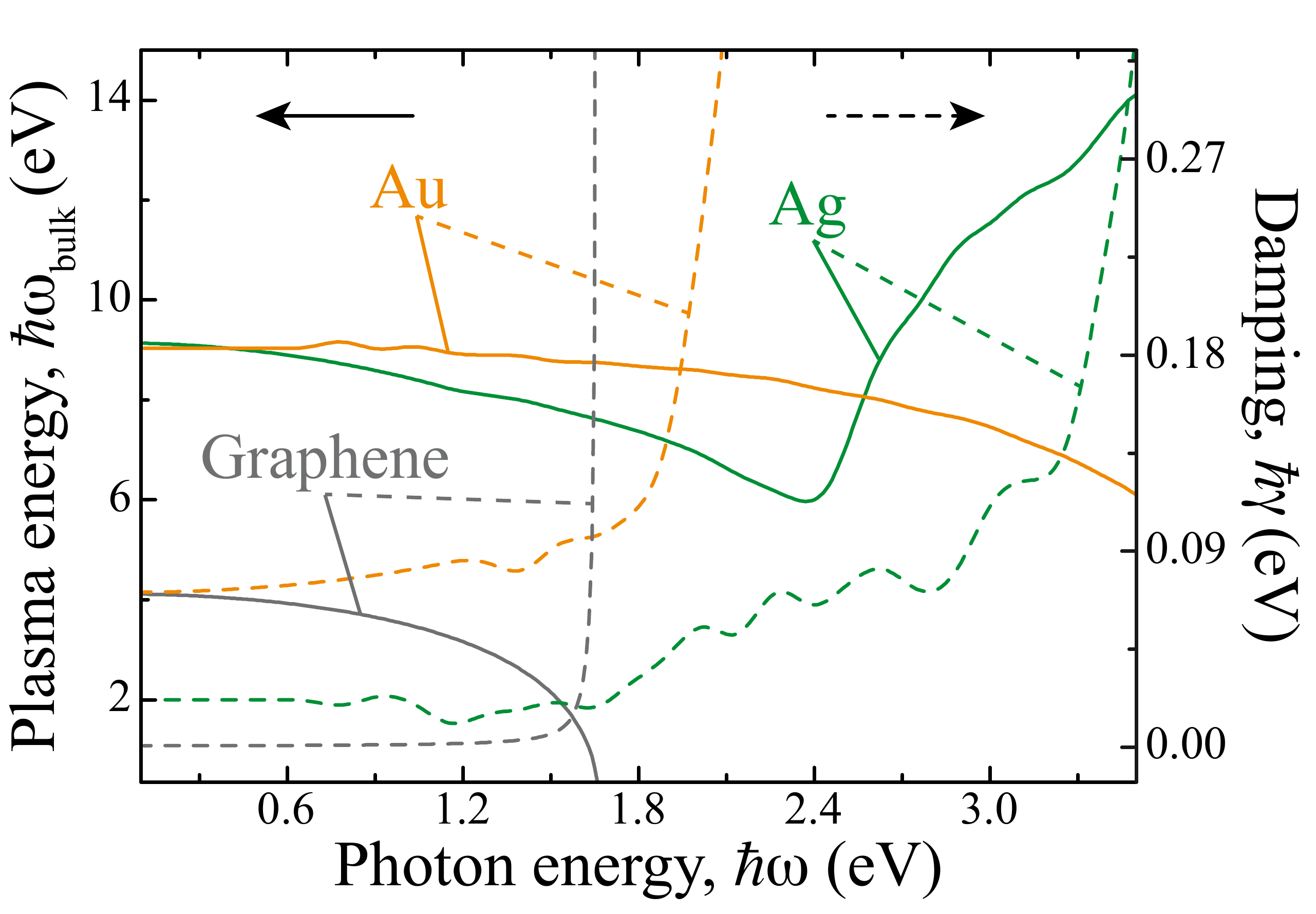}
\caption{{\bf Drude parameters for the response of noble metals.} We represent the $\omega$-dependent parameters $\omegap$ (solid curves) and $\gamma$ for Au and Ag as obtained from $\epsilon(\omega)=1-\omegapp/\omega(\omega+\ii\gamma)$, where $\epsilon(\omega)$ is the measured dielectric function of these materials \cite{JC1972}. The graphene parameters are obtained by fitting Eq.\ (\ref{Drude}) to match the local-RPA conductivity (Eq.\ (\ref{localRPA})) with Fermi energy, mobility, and temperature $E_F=1\,$eV, $\mu=10000\,$cm$^2/(V\,s)$, and $T=300\,$K, respectively, assuming a film thickness $t=0.34\,$nm equal to the interatomic plane distance in graphite. Notice that $\omegap$ is no longer real for $\hbar\omega>1.66\,$eV in graphene (i.e., above the range of plasmonic response for the chosen doping level).} \label{Fig5}
\end{center}
\end{figure}

\section{DRUDE PARAMETERS FOR NOBLE METALS AND GRAPHENE}
\label{Drudegraph}

We show in Fig.\ \ref{Fig5} the Drude parameters $\omegap$ and $\gamma$ for silver, gold, and graphene. For noble metals, we obtain these parameters by fitting the measured dielectric function of the material \cite{JC1972} to the expression $\epsilon(\omega)=1-\omegapp/\omega(\omega+\ii\gamma)$. For graphene, we use the local-RPA conductivity \cite{FV07,paper176}, which we correct in the following expression to simultaneously account for inelastic attenuation and finite temperature $T$ in both intraband and interband transitions \cite{paper235}:
\begin{equation}
\sigma(\omega)=\frac{-e^2}{\pi\hbar^2}\frac{\ii}{\omega+\ii\tau^{-1}}\int_{-\infty}^\infty dE\;\left[|E|\frac{\partial f_E}{\partial E}+\frac{(E/|E|)}{1-4E^2/[\hbar^2(\omega+\ii\tau^{-1})^2]}\;f_E\right],
\label{localRPA}
\end{equation}
where $f_E=1/[1+\ee^{(E-E_F)/k_BT}]$ is the Fermi-Dirac distribution as a function of electron energy $E$ and Fermi energy $E_F$. The first term inside the integral of Eq.\ (\ref{localRPA}), which corresponds to intraband electron-hole pair transitions within the partially occupied Dirac cones of the doped carbon layer, can be integrated analytically to yield a contribution $\ii e^2 E_F^{\rm eff}/\hbar^2(\omega+\ii/\tau)$ with $E_F^{\rm eff}=E_F+2k_BT\,\log\left(1+\ee^{-E_F/k_BT}\right)$. The second term, which originates in interband transitions between lower and upper Dirac cones, needs to be integrated numerically. In Fig.\ \ref{Fig5}, we represent $\omegap$ and $\gamma$ for graphene by fitting Eq.\ (\ref{Drude}) to the values of $\sigma(\omega)$ calculated from Eq.\ (\ref{localRPA}).

\acknowledgments

This work has been supported in part by the European Commission (Graphene Flagship CNECT-ICT-604391 and FP7-ICT-2013-613024-GRASP). A. M. acknowledges financial support from the Welch foundation through the J. Evans Attwell-Welch Postdoctoral Fellowship Program of the Smalley Institute of Rice University (Grant No. L-C-004).


\end{document}